\documentclass[twocolumn]{revtex4}

\usepackage{color} 
\usepackage{graphicx} 
\usepackage{amsmath,amssymb}

\begin{document}

\title{The quest for self-consistency in hydrogen bond definitions}

\author{Diego Prada-Gracia\footnote[1]{these authors contributed equally (in alphabetical order)}}
\author{Roman Shevchuk\footnotemark[1]}
\author{Francesco Rao \footnote[2]{Corresponding author;
E-mail: francesco.rao@frias.uni-freiburg.de; Phone: +49 (0)761 203 97336, Fax:
+49 (0)761-203 97451}} 
\affiliation{Freiburg Institute for Advanced Studies,
School of Soft Matter Research, Albertstrasse 19, 79104 Freiburg im Breisgau,
Germany}

\date{\today}

\begin{abstract}

In the last decades several hydrogen-bond definitions were proposed by
classical computer simulations. Aiming at validating their
self-consistency on a wide range of conditions, here we present a
comparative study of six among the most common hydrogen-bond
definitions for temperatures ranging from 220K to 400K and six
classical water models. Our results show that, in the interval of
temperatures investigated, a generally weak agreement among
definitions is present. Moreover, cutoff choice for geometrically
based definitions depends on both temperature and water model. As
such, analysis of the same water model at different temperatures as
well as different water models at the same temperature would require
the development of specific cutoff values. Interestingly, large
discrepancies were found between two hydrogen-bond definitions which
were recently introduced to improve on more conventional methods. Our
results reinforce the idea that a more universal way to characterize
hydrogen-bonds in classical molecular systems is needed.

\end{abstract}

\maketitle

\section{Introduction}

Hydrogen bonding represents a fundamental interaction in molecular
systems \cite{jeffrey1997}. Its peculiarity resides in the common
aspects it has with both covalent bonds and van der Waals
interactions.  The strong directionality together with the ease of
being formed and broken at ambient conditions makes it an important
ingredient in water structure and dynamics \cite{fecko2003}, protein
stability \cite{mcdonald1994} and ligand binding \cite{wade1993}.
Notwithstanding, a universal definition of this interaction is still
missing \cite{IUPAC}.

Hydrogen-bonds are formed between two polar atoms via a hydrogen which
is covalently bound to one of the two. This interaction is highly
directional. For example, in bulk water at 300K the angle OH-O is
mostly below 30 degrees \cite{Teixeira1990}, while the donor-acceptor
distance is of around 3.5 \AA\ \cite{Soper1986}. Despite the apparent
simplicity, the presence of thermal fluctuations as well as the
non-trivial effects of the environment made the development of an
operative definition of this bond difficult.

In the last decades, several definitions were proposed based on
computer simulations \cite{Matsumoto2007}. The most popular ones look
at bond formation by using a mixture of distances and angles between
the two partners \cite{Kumar2007,Buch1992,Luzar1996}. Others tried to
avoid altogether cutoffs by proposing topology-based definitions
\cite{Hammerich2008,Smith2005,Henchman2010}.  Given the many degrees
of freedom involved in molecular association, it is now clear that all
definitions retain some degree of arbitrariness \cite{agmon2011}.

In most cases, hydrogen-bond definitions were developed at specific
thermodynamic conditions. However, not much is known on the behavior
of those definitions as a function of temperature and water model. In
this research paper, we present a critical assessment of six classical
hydrogen-bond definitions based on the analysis of molecular dynamics
simulations of water in a temperature range spanning from 220K to
400K. Six among the most widespread classical water models were
used in the analysis, including SPC \cite{Berendsen1981}, SPC/E
\cite{Berendsen1987}, TIP3P \cite{Jorgensen1983}, TIP4P
\cite{Jorgensen1985}, TIP4P-Ew \cite{Horn2004} and TIP4P/2005
\cite{Abascal2005}. Our investigation elucidates, in a statistical
way, common features and limitations of current classical
hydrogen-bond definitions.

This work is an effort to present a transparent comparison between hydrogen-bond definitions in several different conditions, including temperature, water model and cutoff dependence. Scientist willing to know the consequences of choosing one method or the other will find our results a good starting point for further development.

\section{Methods}

\begin{figure*}
\includegraphics[width=120mm]{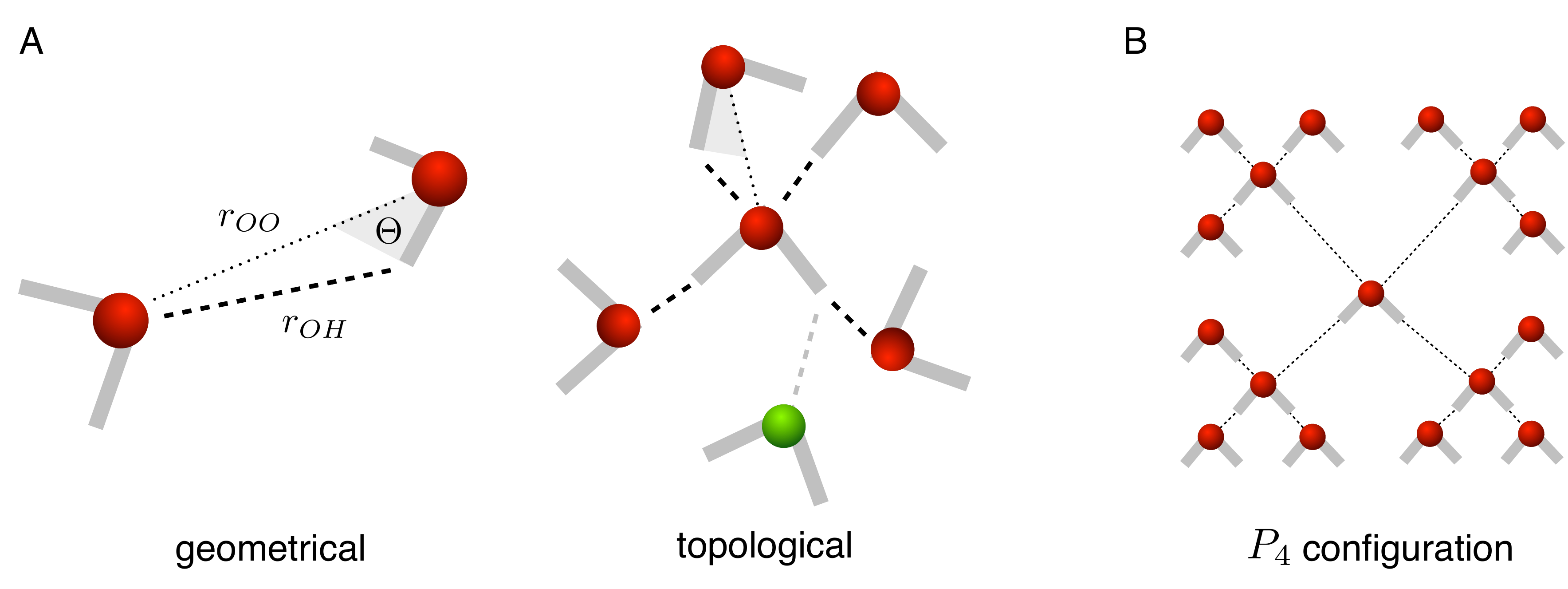}
\caption{(A) Hydrogen-bond definitions can be roughly partitioned into
  two classes: \emph{geometrical} and \emph{topological}. (B) A water
  molecule with a fully-coordinated first and second solvation shells
  is shown as $P_{4}$ configuration, see Methods for details.}
\label{fig:models}
\end{figure*}

\subsection{Hydrogen-bond definitions}

Six hydrogen-bond definitions were considered. They can be classified
into two broad classes: \emph{geometrical} and \emph{topological}
(Fig.~\ref{fig:models}). An important difference between the two is
that geometrical definitions make use of cutoffs on inter-atomic
distances and angles while the latter mostly avoid this problem.  A
brief description of the definitions follows.

\subsubsection*{Geometrical definitions}

\begin{enumerate}

  \item \emph{$r_{OH}$}.  In this definition the oxygen-hydrogen
    distance ($r_{OH}$) is used as criterion (Fig.~\ref{fig:models}A)
    \cite{Buch1992}. In the original work, a cutoff of 2.3 \AA\ was
    proposed by simulating amorphous ice at T=10K with the TIPS2
    potential \cite{Jorgensen1982}. The distance cutoff value is
    related with the position of the first minimum in the
    oxygen-hydrogen radial distribution function.

  \item \emph{$r_{OO}\Theta$}.  This definition makes use of both the
    oxygen-oxygen distance ($r_{OO}$) and the $\angle OOH$ angle
    ($\Theta$) between two water molecules. In the original work, a
    bond was considered formed when $r_{OO}$ and $\Theta$ were smaller
    than 3.5 \AA\ and 30 degrees, respectively~\cite{Luzar1996}. The
    distance cutoff was taken from the position of the first minimum
    in the oxygen-oxygen radial distribution function. Missing a clear
    signature of the bond state in the distribution of the angle
    $\Theta$, the cutoff value was taken from experimental data
    \cite{Soper1986,Teixeira1990}.

  \item \emph{Sk}.  The hydrogen-bond definition of Skinner and
    collaborators is based on an empirical correlation between the
    occupancy $N$ of the $O \cdots H$ $\sigma^{*}$ orbital and the
    geometries observed in molecular dynamics simulations
    \cite{Kumar2007}. Two water molecules were considered bonded if
    the value of $N$ is higher than a certain cutoff which is taken in
    correspondence to the position of the first minimum in the
    distribution of $N$. In the original paper a cutoff equal to
    0.0085 was chosen by analyzing MD simulations of the SPC/E model
    at ambient conditions.

\end{enumerate}

\subsubsection*{Topological definitions}

\begin{enumerate} \setcounter{enumi}{3}

  \item \emph{D$\Theta$}.  A hydrogen-bond is formed between a
    hydrogen atom and its nearest oxygen not covalently bound. An
    additional restriction was imposed: the angle $\Theta$ had to be
    lower than $\pi /3$. In the original work \cite{Smith2005}, this
    definition was applied to the study of the SPC/E water model for
    temperatures ranging from 273 to 373K.

  \item \emph{DA}.  Two criteria for the hydrogen-bond were used: (i)
    the acceptor is defined as the closest oxygen to a donating
    hydrogen and (ii) this hydrogen is the first or second nearest
    neighbor of the oxygen. As a consequence, the total number of
    hydrogen-bonds per water is limited to four. This definition was
    proposed with simulations of the EMP water model at 292K
    \cite{Hammerich2008}.

  \item \emph{TP}. A hydrogen-bond is formed between a hydrogen and
    its closest oxygen. When more than one hydrogen-bond between the
    two water molecules is found, the one with the shortest
    oxygen-hydrogen distance is considered to be formed
    \cite{Henchman2010}. This definition was mainly evaluated at
    ambient conditions using the TIP4P/2005 water model.

\end{enumerate}

The six hydrogen-bond definitions above as well as all the analysis
tools applied in this work were implemented in the software {\sc
  aqua-lab} freely distributed at {\tt raolab.com}.

\subsection{Water structural propensities including the second shell}

Hydrogen-bond propensities up to the second solvation shell were
obtained by calculating the following parameters: the probability
$P_{4}$ to have a water molecule with a fully-coordinated first and
second solvation shells, for a total of 16 bonds (right panel of
Fig.~\ref{fig:models}); and the probability to have four ($P_4^{*}$),
three ($P_3$) and two or less ($P_{210}$) bonds with a generic first
solvation shell. In the calculation of $P_4^{*}$ the propensity of
$P_{4}$ was subtracted. A more comprehensive study of these four
propensities, including temperature and water model dependence, was
already presented elsewhere \cite{Shevchuk2012}.

\subsection{Hydrogen-bond kinetics}

Kinetics was analyzed in terms of hydrogen-bond lifetime distributions. The lifetime was calculated as follows. For each definition, pairwise hydrogen bonds among all water molecules were calculated for every frame. For each of the water pairs that formed a bond, the time span for how long that particular bond lasted is called lifetime. The distribution was then calculated by building an histogram of all the lifetimes collected in the molecular trajectory. The average lifetime was denoted with the symbol $\tau$.

\subsection{Molecular dynamics simulations}

All molecular dynamics simulations (MD) were run with GROMACS
\cite{gromacs}. The setup included a cubic water box of 1024 water
molecules. Temperature and pressure were controlled with the velocity
rescale thermostat \cite{Bussi2007} and Berendsen barostat
\cite{Berendsen1984}, respectively. In both cases a 1 ps relaxation
time was used. Non-covalent interactions were treated with a 1 nm
cutoff and Particle-Mesh-Ewald \cite{Darden1993}. Each water model was
simulated for 1 ns after a 10 ns equilibration run, saving coordinates
every 4 fs for a total of 250'000 snapshots per model.

For the present analysis six popular classical water models, namely
SPC \cite{Berendsen1981}, SPC/E \cite{Berendsen1987}, TIP3P
\cite{Jorgensen1983}, TIP4P \cite{Jorgensen1985}, TIP4P-Ew
\cite{Horn2004} and TIP4P/2005 \cite{Abascal2005} were simulated for
temperatures ranging from 220K to 400K at ambient pressure.

\section{Results}

\subsection{Hydrogen-bond propensities in temperature space}

\begin{figure}
\includegraphics[width=80mm]{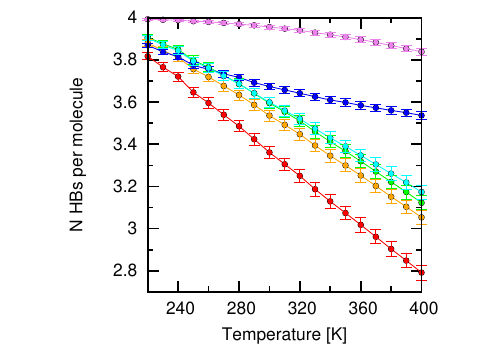}
\caption{Average number of hydrogen bonds per water molecule for the
  six hydrogen-bond definitions: $r_{OH}$ (orange), $r_{OO}\Theta$
  (green), $Sk$ (red), $D\Theta$ (cyan), $DA$ (blue) and $TP$
  (purple)}
\label{fig:num_hbs}
\end{figure}

MD simulations of 1024 SPC/E water molecules were performed for
temperatures ranging from 220K to 400K at ambient pressure.  For
each condition the average number of hydrogen-bonds per molecule was
calculated.  Results for the six definitions under study (see Methods
for details) are shown in Fig.~\ref{fig:num_hbs}. As expected, the
general trend shows a decreasing number of hydrogen-bonds per molecule
by increasing temperature. At 220K this number was very similar among
all approaches ranging from 3.8 to 4.0.  However, at ambient
temperature definitions deviated quite considerably. $Sk$ provided the
smallest number of bonds (3.37 per molecule), while, with an average
connectivity of 3.95, the purely topological definition ($TP$)
provided the largest number. This discrepancy amplified as the
temperature was further increased to 400K with values
approaching 2.79 and 3.84 for $Sk$ and $TP$,
respectively. Better agreement along the whole temperature range was
found among $r_{OH}$, $r_{OO}\Theta$ and $D\Theta$. It is interesting
to note that among these definitions, $Sk$ was optimized against
ab-initio calculations using the SPC/E model at 300K (same model as
here). At least in principle, Skinner's data should provide the most
accurate estimation at 300K and ambient pressure.

\begin{figure}
  \includegraphics[width=50mm,angle=270]{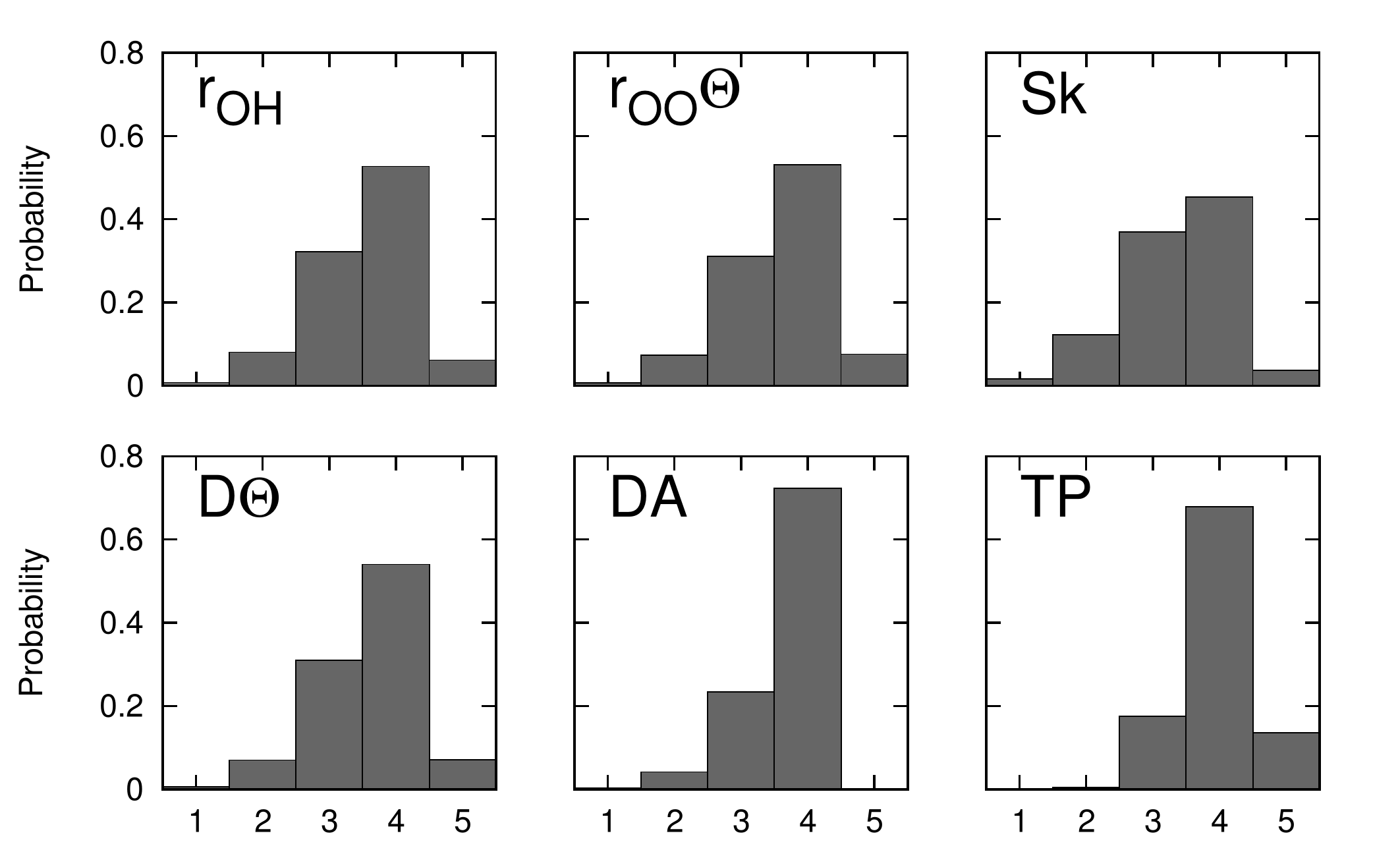}
  \caption{Average number of bonded partners for the six hydrogen-bond definitions at 300K.}
  \label{fig:nhbs}
\end{figure}

Discrepancies were also found in the distribution of the number of
bonded partners (Fig.~\ref{fig:nhbs}). At 300K geometrical definitions
were quite consistent among each other, with a larger fraction of
three hydrogen-bonded configurations for $Sk$. For the topological
case, $DA$ and $TP$ agreed on the number of four coordinated
molecules.  However, the former detected a larger fraction of three
and two bonded molecules while $TP$ presented a non-negligible
fraction of cases with five partners and no evidence for two bonded
molecules. This scenario changes when a topological definition is
coupled with an angle cutoff ($D\Theta$).  In this case, almost
identical results as the conventional $r_{OO}\Theta$ were found with
an agreement that persists in the entire temperature range as shown in
Fig.~\ref{fig:num_hbs} (green and cyan data).

\begin{figure*}
  \includegraphics[width=140mm]{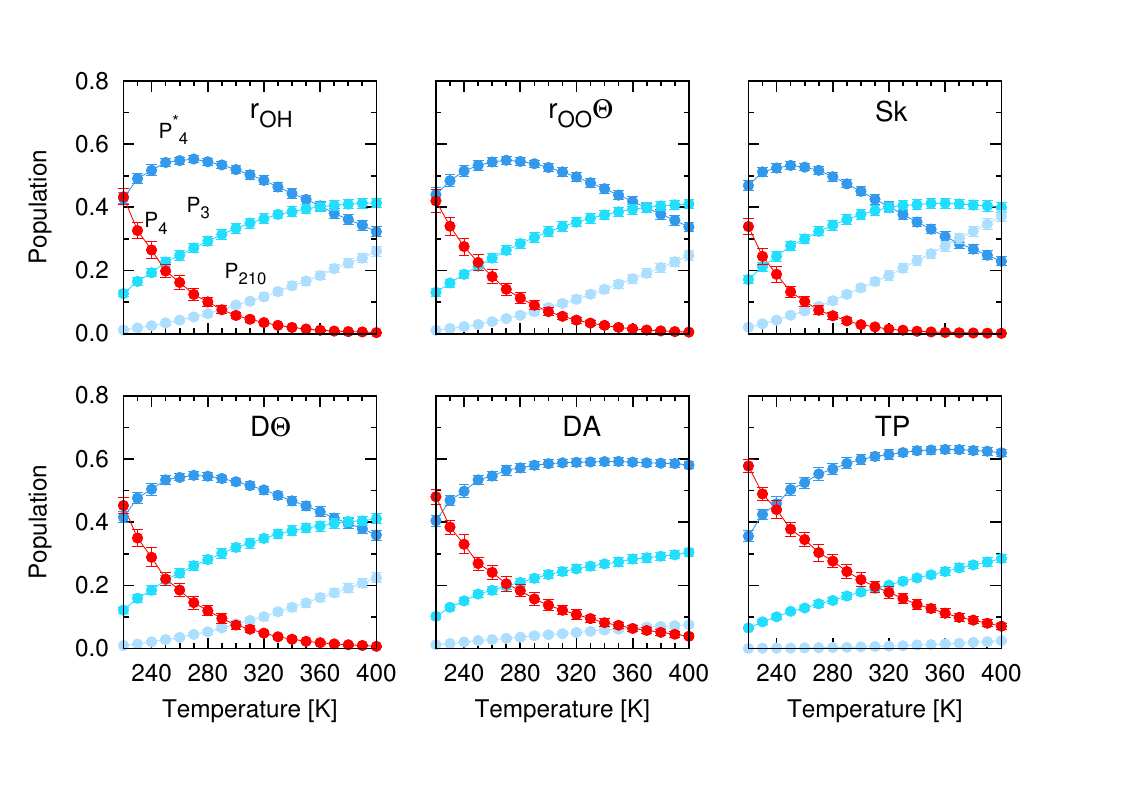}
  \caption{Hydrogen-bond propensities including the second solvation
    shell for temperatures between 220K-400K. $P_4$,
    $P_{4}^{*}$, $P_3$ and $P_{210}$ are shown in red, blue, light
    blue and very light blue, respectively (see Methods for details).}
  \label{fig:pi}
\end{figure*}

In Fig.~\ref{fig:pi} hydrogen-bond propensities including the second
solvation shell are presented.  The behavior of these propensities
strongly depend on the hydrogen-bond definition taken into
account. Consistency was found within two groups. The first one
includes $r_{OH}$, $r_{OO}\Theta$ and $D\Theta$ and the second one
$TP$ and $DA$. $Sk$ did not match very well any of them.  The value of
$P_4$, i.e., the probability to have a four-coordinated water molecule
with a fully coordinated first and second shells (Methods and
Fig.~\ref{fig:models}B), was equal to 0.34 and 0.58 at 220K for $Sk$
and $TP$, respectively (red data).  As temperature was increased this
difference became even more pronounced.  A similar disagreement was
also observed for the other three propensities.

An interesting case is given by $P_4^*$. This quantity reports on
four-coordinated water molecules with an arbitrarily disordered second
solvation shell. For all definitions this quantity presented a
  peak.  However, $TP$ and $DA$ made an exception being the maximum
  much more shallow and at a higher temperature with respect to the
  other approaches. This leads to an over estimation of four
  coordinated water molecules which are predicted to be the most
  abundant configuration at temperatures as high as 400K.  This result
  is counter intuitive as waters with three or less hydrogen-bonds
  would have been expected to represent a larger fraction of the
  sample at a such high temperature. Substantial discrepancies among
definitions were also found in the case of $P_{210}$ (water molecules
with two bonds or less). For the case of $TP$ this probability was
essentially zero at all temperatures while it grew with temperature in
all the other cases.

\subsection{Hydrogen-bond lifetime}

\begin{figure}
  \includegraphics[width=80mm]{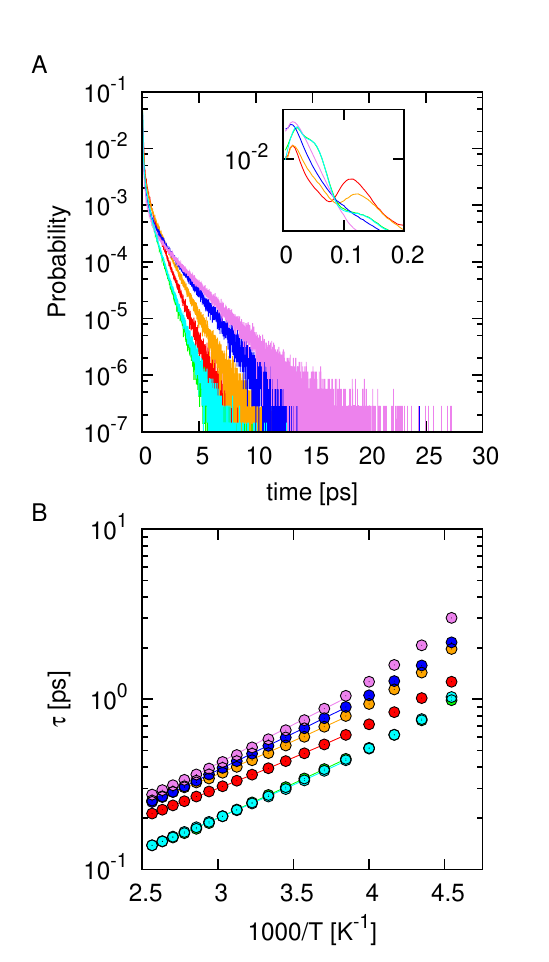}
  \caption{Hydrogen-bond kinetics for the six different definitions:
    color-code is the same as in Fig. 2. (A) The lifetime distribution
    at T=300K is shown. (B) the average hydrogen-bond lifetime versus
    1/T is plotted (error bars are smaller than the symbol size). The
    Arrhenius behavior is observed in the range of temperatures from
    260 to 400K.}
  \label{fig:fpt}
\end{figure}

Water kinetics was analyzed through hydrogen-bond lifetimes (see
  Methods). Distributions for the six hydrogen-bond definitions at
300K are shown in Fig.~\ref{fig:fpt}A. Fastest decays (i.e. shorter
life times) were observed for $r_{OO}\Theta$ (green) and $D\Theta$
(cyan), strongly suggesting that fluctuations along the $\Theta$ angle
represent the major responsible for the faster kinetics. On the other
hand, the largest lifetimes were found with the $TP$ definition. At
very short times ($<200fs$) purely topological approaches provided the
best results (inset of Fig.~\ref{fig:fpt}A). In fact, both $DA$ (blue)
and $TP$ (purple) showed a smooth decay, in contrast to all the other
definitions which provided a debatable oscillating behavior
\cite{Starr1999}. This observation strongly suggests that those
fluctuations are an artifact of the use of cutoffs.

For the average lifetime $\tau$, Arrhenius behavior in the range
260K$<$T$<$400K was found, breaking down for lower temperatures
(Fig.~\ref{fig:fpt}B). $r_{OO}\Theta$, $D\Theta$ and $r_{OH}$, $DA$,
$TP$ provided fastest and slowest timescales, respectively. $Sk$
shifted from one group to the other while changing temperature (red
data in Fig.~\ref{fig:fpt}B).

\subsection{Self-consistency}

\begin{figure*}
\includegraphics[angle=0,width=0.7\textwidth]{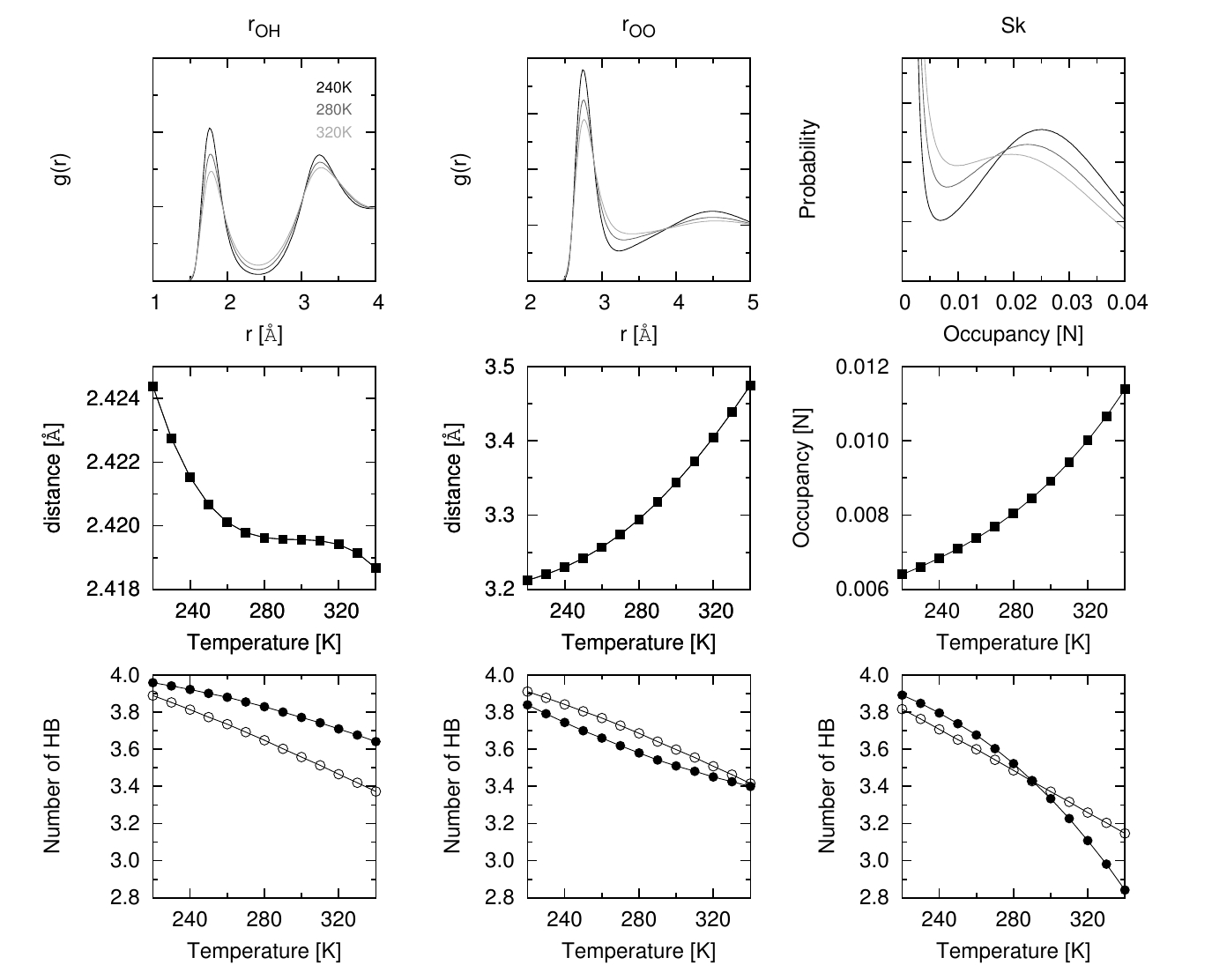}
\caption{Temperature dependence for cutoff choice. Data relative to
  the $r_{OH}$, $r_{OO}\Theta$ and $Sk$ definitions are shown in the
  first, second and third column, respectively. (Top) The
  oxygen-hydrogen, oxygen-oxygen radial distribution functions and the
  occupancy distribution are displayed from left to right. (Middle)
  Cutoff dependence as a function of temperature. (Bottom) Average
  number of hydrogen bonds with fixed and variable cutoffs are shown
  as empty and filled circles, respectively.}
\label{fig:cutoff}
\end{figure*}

Hydrogen-bonds were described so far on the base of propensities and
kinetics. Now, we investigate the robustness of the geometrical
definitions with cutoff choice.  The aim of the following analysis is
to understand what is the influence of temperature and water model on
the distributions relevant to cutoff choice.  In fact, default cutoff
values were originally proposed from experiments and calculations at
specific temperatures and water models (see Methods). Although in most
cases prescriptions were given to properly choose the cutoffs, default
values were often applied in conditions far away from the original
works.

First, temperature dependence was investigated. For $r_{OH}$, the
distribution that matters is the oxygen-hydrogen radial distribution
function ($g(r)$, left column in Fig.~\ref{fig:cutoff}). The plot
shows that the first minimum becomes less pronounced with
temperature while its position gets closer to the origin (from 2.424
to 2.419 \AA, left middle row of Fig.~\ref{fig:cutoff}).  Choosing the
cutoff according to the position of the minimum, the average number of
hydrogen-bonds per molecule was significantly affected despite the
small change of the cutoff value. In the bottom left panel of
Fig.~\ref{fig:cutoff} the difference between a ``standard'' cutoff
approach (empty circles) and a temperature dependent cutoff (filled
circles) is shown.

Similar results were obtained for the other two geometrical
definitions. For these cases the value of the cutoff was chosen
according to the position of the first minimum of the oxygen-oxygen
radial distribution function and the distribution of the occupancy $N$
for the case of $r_{OO}\Theta$ and $Sk$, respectively (second and
third columns of Fig.~\ref{fig:cutoff}).  Lacking of a bimodal
behavior we intentionally avoided the study of the angle $\Theta$
cutoff dependence.

\begin{figure}
\includegraphics[angle=0,width=0.5\textwidth]{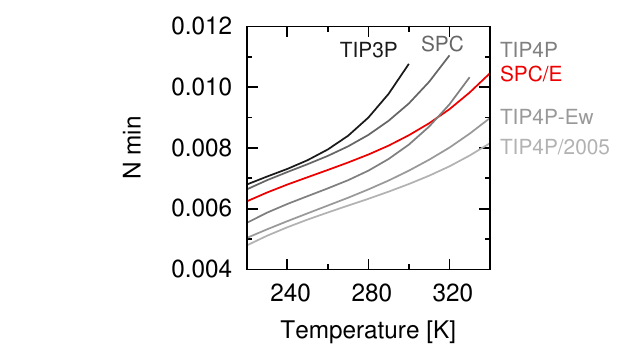}
\caption{The position of the first minimum of the occupancy
  distribution relative to the $Sk$ definition for different water
  models. Red points refer to the SPC/E model which was used for the
  rest of the analysis presented in this paper.}
\label{fig:skinner}
\end{figure}

Interestingly, radial distribution functions depend not only on
temperature but also on the water model under study. This suggests a
further dependence on cutoff choice. To verify this idea, we ran MD
simulations of six of the most commonly used water models (see Methods
for details). In Fig.~\ref{fig:skinner} results for the $Sk$
definition are presented. The data reports on the position of the
first minimum in the occupancy $N$ distribution as a function of
temperature for different water models. According to the original
prescription \cite{Kumar2007}, the hydrogen-bond cutoff should be
taken as the position of this minimum. The plot shows that this value
strongly depends on both water model and temperature. Similar
conclusions can be drawn for the case of $r_{OH}$ and $r_{OO}$.

\section{Discussion}

Despite the fundamental role of hydrogen bonding in molecular
processes, a robust, self-consistent and universally accepted
definition is still missing. In the context of classical molecular
simulations several proposals emerged in recent years. In this work,
we systematically analyzed six among the most popular of these
definitions, trying to elucidate their range of validity, strengths
and limitations.  The definitions under study can be classified into
two main groups: \emph{geometrical} and \emph{topological}.

Geometry-based methods make use of cutoffs. The latter identify a
geometrical volume where the hydrogen-bond is thought to be formed.
However, the numerical value of a cutoff is usually chosen with some
degree of arbitrariness.  We found that the distributions behind the
definitions of a cutoff are strongly dependent on temperature and
water model.  As a consequence, hydrogen-bond cutoffs would need
careful evaluation for each temperature and water model in order to be
self-consistent with the underlying distributions. Unfortunately, the
use of optimal cutoffs do not necessarily lead to improved consensus
among hydrogen-bond definitions. The cutoff optimization presented
here is similar in spirit to an interesting prescription recently
suggested by Skinner and co-workers \cite{Kumar2007}.  In that case,
hydrogen-bond boundaries were defined by the isoline passing through
the saddle point of bond forming/breaking of the corresponding
coordinates taken into account (e.g. $R_{OO}$ and $\Theta$).  This
approach is certainly optimal when the free-energy projection onto
those coordinates represents an accurate description of the system
kinetics. However, given the large number of degrees of freedom
involved, this assumption does not hold in general \cite{Geissler1999,
  Rao2004, Krivov2004,Rao2010, berezovska2012}.  Looking at the
re-crossings present on top of the barrier for hydrogen-bond
forming/breaking as well as the oscillations observed in the lifetime
distribution at short times suggest that those maps do not describe
the kinetics well.

To cure some of the issues related to a cutoff based approach,
topology-based definitions were recently introduced
\cite{Hammerich2008, Henchman2010}.  However, these approaches are
characterized by a very small number of broken hydrogen bonds at high
temperatures: a trend that seems to be debatable.

Overall, our analysis put in evidence a number of limitations in
current approaches, highlighting a general lack of consensus among
them. Somewhat surprising was to find that two of the most recent
definitions, $Sk$ and $TP$, were the ones to agree the least with each
other. This certainly motivates the exploration of alternative routes,
like the use of multi-body definitions going beyond the classical
pairwise models \cite{Markovitch2007, Znamenskiy2007}.  However, we
speculate that the most overlooked player in this game might very well
be \emph{kinetics}. Our suggestion is motivated by the recent
approaches emerged in the study of protein conformational changes
which constructively used system kinetics to characterize the
underlying free-energy landscape. Based on
configuration-space-networks \cite{Rao2004,Krivov2004} and
Markov-state-models \cite{swope2004,Chodera2007,berezovska2012} those
methods provided projection-free state definitions and interconversion
rates.  As such, two structures are detected to belong to the same
\emph{state} if their dynamics is similar, shifting the problem from
being geometrically similar to be kinetically similar.  Applications
of these ideas to the study of bulk water were recently put forward by
our group \cite{Rao2010, Prada2012} but a straightforward extension to
hydrogen-bond structure and dynamics is not trivial.  Given the
fundamental role of hydrogen bonding in molecular systems, we believe
that such type of developments will obtain increased attention in the
future.

\section{Acknowledgment}

This work is supported by the Excellence Initiative of the German
Federal and State Governments.


\end{document}